\documentclass[journal]{IEEEtran}
\usepackage{generic}
\usepackage{cite}
\usepackage{amsmath,amssymb,amsfonts}
\usepackage{algorithmic}
\usepackage{graphicx}
\usepackage{textcomp}
\usepackage{subcaption}
\usepackage{comment}
\usepackage{array}
\usepackage{hhline}
\usepackage{float}
\usepackage{siunitx}

\usepackage[english]{babel}

\usepackage[colorlinks=true, allcolors=blue]{hyperref}
\usepackage{caption}

\title{Wideband Glide-Symmetric Double-Corrugated Gap-Waveguide Traveling-Wave Tube for Millimeter Waves}

\author{Miguel Saavedra-Melo (\IEEEmembership{Student Member, IEEE}), Nelson Castro (\IEEEmembership{Student Member, IEEE}), Robert Marosi (\IEEEmembership{Member, IEEE}), Eva Rajo-Iglesias (\IEEEmembership{Senior Member, IEEE}), Filippo Capolino (\IEEEmembership{Fellow, IEEE})
\thanks{This material is based upon work partly supported by the Air Force Office of Scientific Research Multidisciplinary Research Program of the University Research Initiative (MURI) under grant number FA9550-20-1-0409 administered through the University of New Mexico. The work was partially funded by the Spanish Government under the grant PID2022-141055NB-C22 from MCIN/AEI/10.13039/501100011033 and in part by RED2022-134657-T.}
\thanks{Miguel Saavedra-Melo, Robert Marosi, and Filippo Capolino are with the Department of Electrical Engineering and Computer Science, University of California, Irvine, CA 92697 USA (e-mail: f.capolino@uci.edu).}
\thanks{Nelson Castro, Eva Rajo Iglesias, and Filippo Capolino and are with the Department of Signal Theory and Communications, University Carlos III of Madrid, Leganés, 28911, Spain (email: ncastro@pa.uc3m.es).}}

\begin{document}

\maketitle

\begin{abstract}
We explore the use of glide symmetry (GS) and electromagnetic bandgap (EBG) technology in a glide-symmetric double corrugated gap waveguide (GSDC-GW) slow wave structure (SWS) for traveling wave tube (TWT) applications. Notably, this GS structure provides the advantage of wide-band operation and the EBG eliminates the need for a conductive connection between the top and bottom waveguide plates.  The TWT performance is evaluated via particle-in-cell simulations that reveal a 3-dB bandwidth of approximately 12 GHz spanning from 54.5 GHz to 66.3 GHz, accompanied by a maximum gain of 23 dB.  Because of GS, the backward wave in the first spatial harmonic is not longitudinally polarized, leading to a low risk of backward wave oscillations in the TWT. This work places the GSDC-EBG structure within the arena of potential SWS topologies for TWTs operating under similar conditions.
\end{abstract}

\section{Introduction}

Traveling wave tubes (TWT) offer several advantages over their solid state power amplifier counterparts that are commonly used in similar applications. These advantages include a moderate dc power consumption, broad bandwidth, high efficiency in transforming dc beam power into amplified radio frequency waves, extensive lifetime, and exceptional reliability even under high thermal stresses and ionizing radiation conditions typical of space and airborne applications \cite{paoloni15,Chong10}. Broadband TWTs have been widely studied, as in the research by Shin, et.al. \cite{shin09}, which demonstrated a half-period staggered double-vane structure capable of an output power ranging from 150 to 270 W while operating within a 50-GHz band centered at 220 GHz, with a 5 kW electron beam. Broadband TWTs were also discussed in ref. \cite{paoloni14}, where a double corrugated waveguide for operation in the G-band (110-300 GHz) was proposed, offering 18 dB gain and output power of up to 3.7 W while covering a bandwidth of approximately 30 GHz at a central frequency of 225 GHz, with 390 W electron beam. Most commercial TWTs use helical slow-wave structures (SWSs), offering broad bandwidths and power outputs exceeding 100 W up to 50 GHz \cite{paoloni15}. Nevertheless, their fabrication becomes challenging and impractical for millimeter-wave frequencies \cite{paoloni15}. In the terahertz (THz) and sub-THz frequency bands, the most common SWS geometry is the serpentine waveguide \cite{Nguyen2014,liu2016characteristics,Liao15V_band_TPS}, capable of providing some of the widest bandwidths for non-helix TWTs with a moderate power output. 


Various interaction structures suitable for millimeter-wave TWTs have been conceived, including the folded waveguide, double staggered waveguide, and double corrugated waveguide, as described in \cite{paoloni21}. In recent years, straight rectangular waveguides loaded with pillars have been introduced as SWSs due to their easier manufacturing, with a simplified beam tunnel integration. For example, the single corrugated waveguide \cite{joye13}, the half-period staggered double vane structures \cite{shin09,shin11}, and the double corrugated rectangular SWS \cite{paoloni15,paoloni14,paoloni13}.

At the same time, gap waveguides (GW) have become popular as a low-cost and low-loss solution for millimeter-wave frequencies \cite{kildal09,Rajo-Iglesias10,Letizia15,Abozied23Half}. They are derived from electromagnetic bandgap (EBG) technology initially explored in \cite{Sievenpiper99} and \cite{Smirnova02}. 
These EBG waveguides are obtained by the substitution of conventional metal walls in a rectangular waveguide with an array of pillars forming an EBG, eliminating mechanical joints where electric currents could flow and reducing resistive losses \cite{kildal09}. Additionally, the GW technology allows for flexible design integration and offers a frequency region where electromagnetic propagation is prohibited, and as a consequence, undesirable higher-order modes in the waveguide can be effectively suppressed \cite{Abozied23Half}.




\begin{figure}[htpb]
\centering
\includegraphics[width=0.94\linewidth]{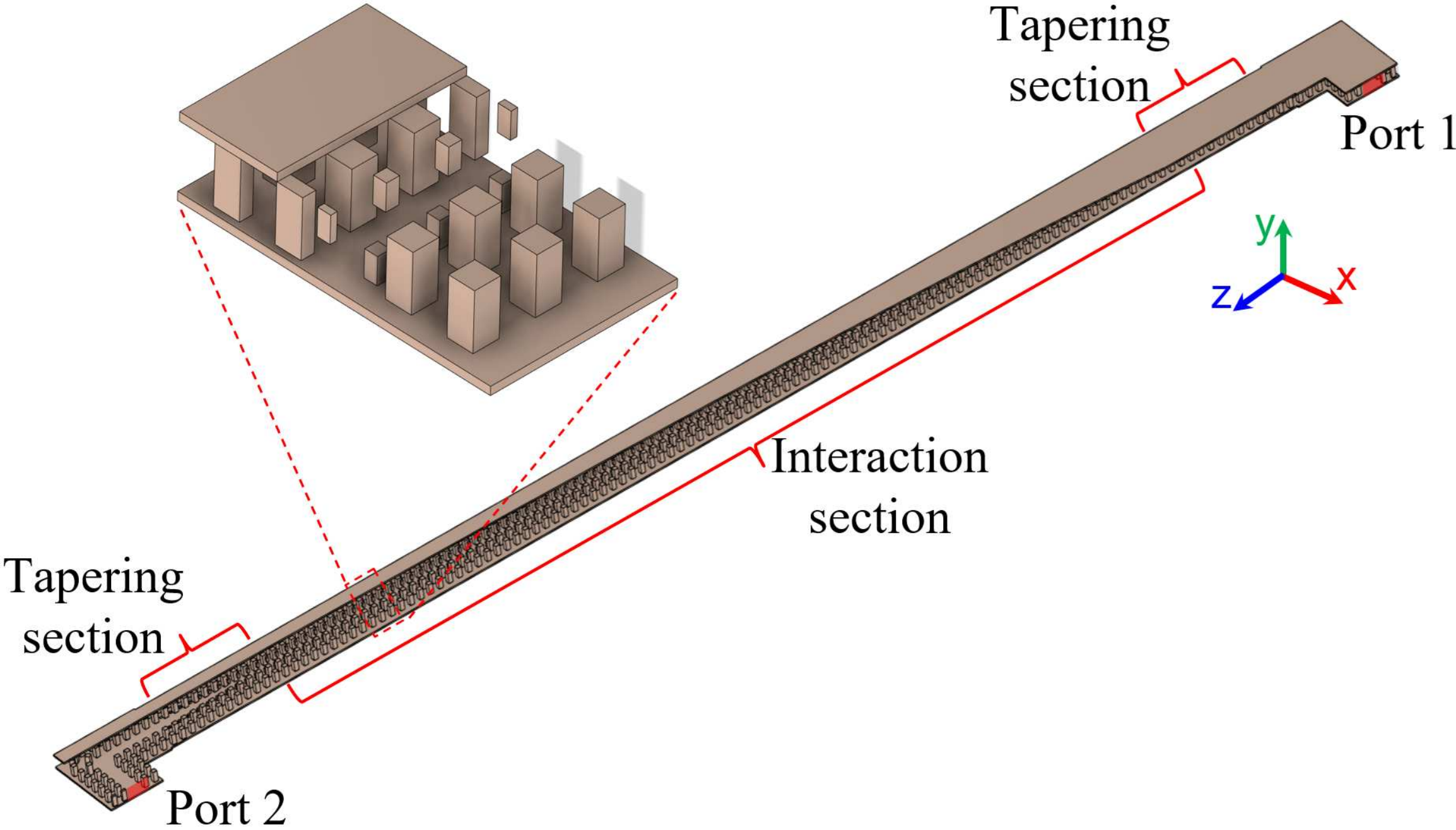}
\caption{Longitudinal cross-section of the SWS for the GSDC-GW TWT, which consists of the interaction, tapering, and bend sections. Each section is glide symmetric for wide band operation.} 
\label{fig:full_str}
\end{figure}

Glide symmetry (GS) is a type of higher symmetry found in periodic structures, achieved through a translation of half the geometrical period along the waveguide's longitudinal direction $z$, followed by a reflection across a glide plane, as detailed in \cite{Hessel73}. The waveguide investigated in this work exhibits GS along not one, but two glide planes. Hence, the GS operation is defined as 
 
\begin{equation}
  G \equiv  \begin{cases}
    x \rightarrow -x \\
    y \rightarrow -y \\
    z \rightarrow z + d/2
    \end{cases}
    \label{eq:GSoperator}
\end{equation}

where $d$ is the period of the SWS. In this paper, the mirror symmetry operation is applied with respect both the $x-z$ plane and $y-z$ plane. Originally explored in \cite{Hessel73,crepeau64,Mittra65}, GS is important due to its remarkable characteristics such as low dispersion (i.e. minor fluctuations in the phase velocity, $v_{\rm{ph}} = \omega / \beta_{\rm{c}}$, throughout the operating bandwidth) and the possibility of closing stopbands at $\beta_{\rm{c}} d=\pi+2\pi n$ points, where $\beta_{\rm{c}}$ is the ``cold" (i.e., without the electron beam present) propagation constant of the guided mode, and $n=0, \pm1, \pm2, \ldots$. Recently, GS has been studied for various applications  \cite{quevedo20, quevedo21,ebrahimpouri18,monje20}. Additionally, in the context of millimeter-wave frequencies, \cite{fischer23, Tan23} describe applications of GS in metasurface waveguides and EBG structures. 
In vacuum electronics, an offset double corrugated SWS with GS corrugations for powering long-range sub-THz links was studied in \cite{Basu2021,Patent}.




Here, we investigate the role of GS in a TWT amplifier using a glide-symmetric double corrugated gap waveguide (GSDC-WG), building upon prior research in \cite{Castro23}. We employ particle-in-cell (PIC) simulations to study the TWT's operational characteristics. The TWT consists of a GS interaction region, tapering sections, and bend sections. We show the GS-induced wide gain bandwidth of the TWT amplifier and discuss its performance that is compared with that of other TWT geometries operating under similar conditions.

\section{SWS With Glide Symmetry }

\subsection{Geometry and Modal Dispersion}


The GS-TWT structure is presented in Fig. \ref{fig:full_str}. The interaction section at the center of the TWT has a length of 180.5 mm, or $N_c=95$ unit cells of period $d$. Additionally, each tapering section features a nonlinear tapering transition consisting of 11 cells, designed to interface with the standard WR-15 rectangular waveguide (3.7592 mm $\times$ 1.8796 mm) at the end of the bends (details about the scattering parameters of each TWT section are in the Appendix).

\begin{figure}[htpb]
\centering
\includegraphics[width=0.96\linewidth]{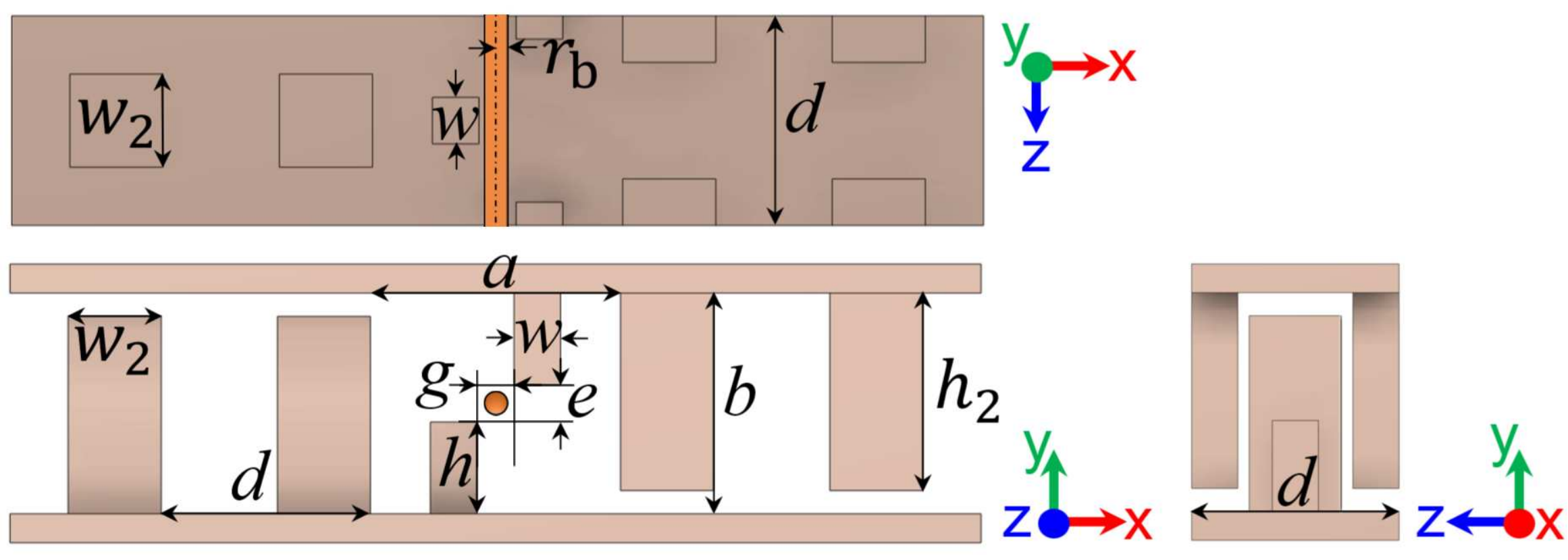}
\caption{Unit cell of the waveguide with glide symmetry as in (\ref{eq:GSoperator}). A rectangular waveguide between contactless metallic plates with periodic pillars of square cross-section inside. The small pillars guide the wave and the large pillars form the EBG. The location of the electron beam is illustrated in orange.}
\label{fig:unit_cell}
\end{figure}

The unit cell's geometry, as illustrated in Fig. \ref{fig:unit_cell}, is a modified version of the bottom-top glide design (BT glide) introduced in \cite{Castro23}, where the vertical side walls have been replaced by EBG pillars. It is inspired by the groove gap waveguide design outlined in \cite{Rajo-Iglesias10} and the offset double corrugated waveguides (DCW) described in \cite{Basu2021} and \cite{Patent}. This SWS incorporates two different types of pillars: two small inner pillars per unit cell within the waveguide core, strategically positioned for optimal interaction impedance enhancement \cite{mineo10} (their proximity, while maximizing interaction, requires careful consideration to minimize electron interception when the beam is present \cite{Basu2018}), and four larger outer pillars that are used to confine the electromagnetic mode in the waveguide with an effective width $a$. These outer pillars are specifically engineered to operate within the stop-band regime and confine the desired waveguide modes \cite{Rajo-Iglesias10}. This feature streamlines the manufacturing and assembly process by eliminating the need for low-loss electrical contact between the top and bottom plates that form the waveguide while allowing wideband operation. Electrical connections may still occur between the two metal plates beyond the outer pillars to maintain the required vacuum environment but there, surface currents do not need to be supported due to the EBG design.

The geometry in Fig. \ref{fig:unit_cell} has been optimized to allow for synchronization between the electromagnetic mode and the electron beam with a circular cross-section of a radius, $r_{\rm{b}} = 0.1$ mm, and an average charge velocity of $u_0=0.283c$ (corresponding to an accelerating voltage $V_0= 21.8$ kV), where $c$ is the speed of light. The dimensions are in Table \ref{tab:dims}.

\begin{table}[H]
\caption{Dimensions of the GSDC-GW unit cell in the interaction section}
\centering
\begin{tabular}{|>{\centering\arraybackslash}p{1.3cm}|>{\centering\arraybackslash}p{0.3cm}|>{\centering\arraybackslash}p{0.3cm}|>{\centering\arraybackslash}p{0.3cm}|>{\centering\arraybackslash}p{0.3cm}|>{\centering\arraybackslash}p{0.3cm}|>{\centering\arraybackslash}p{0.3cm}|>{\centering\arraybackslash}p{0.3cm}|>{\centering\arraybackslash}p{0.3cm}|>{\centering\arraybackslash}p{0.3cm}|}
\hline
\textbf{Dimension} & \textbf{$a$} & \textbf{$b$} & \textbf{$d$} & \textbf{$e$} & \textbf{$g$} & \textbf{$h$} & \textbf{$h_2$} & \textbf{$w$} & \textbf{$w_2$} \\ 
\hline
\textbf{Val. (mm)} & 2.05 & 1.9 & 1.9 & 0.32 & 0.32 & 0.79 & 1.7 & 0.4 & 0.8 \\ 
\hline
\end{tabular}
\label{tab:dims}
\end{table}

\begin{figure}[htpb]
\centering
 \includegraphics[width=0.9\linewidth]{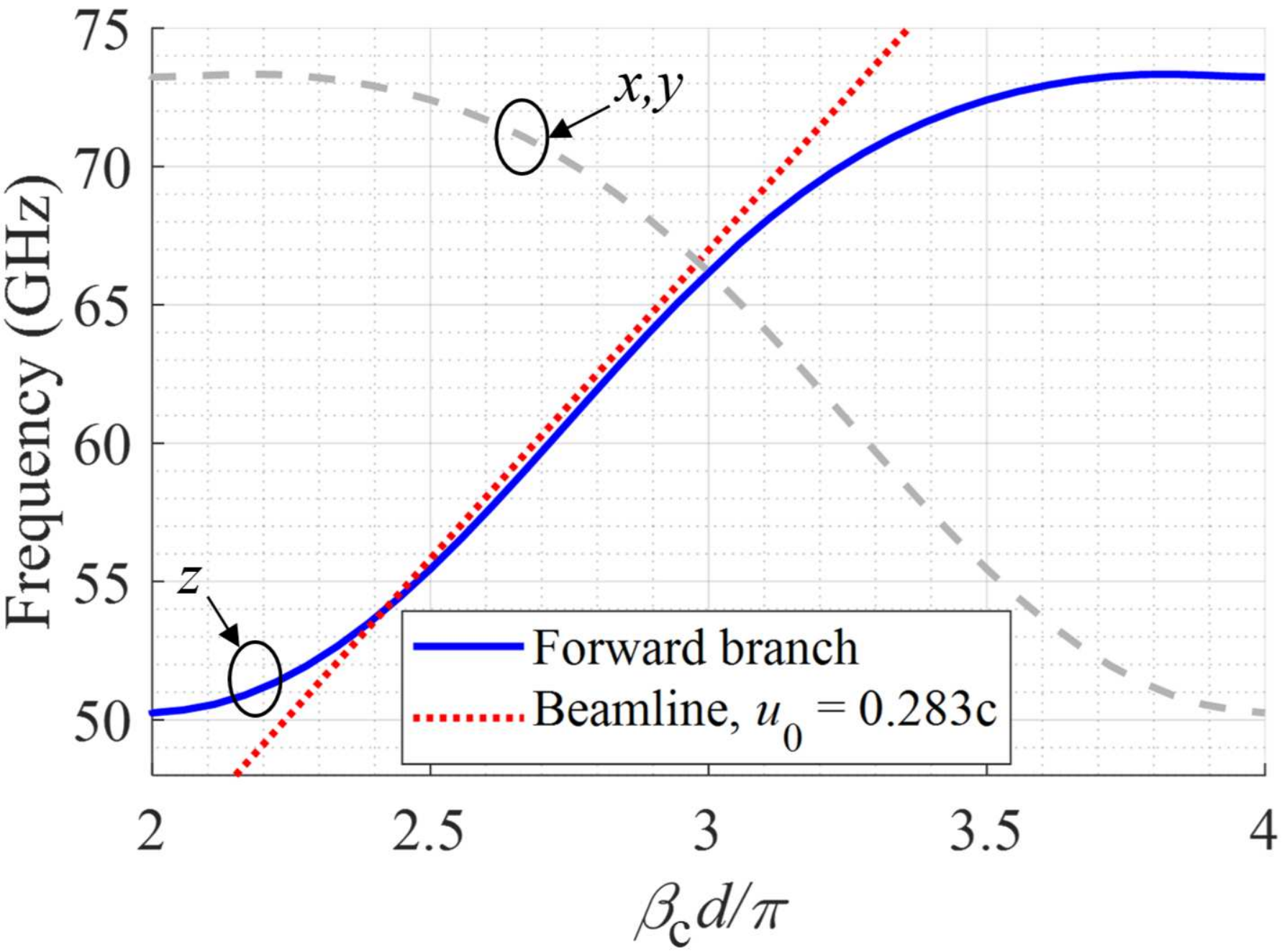}
\caption{Modal dispersion diagram and polarization of the electric field in the cold SWS: harmonic with the positive slope in solid blue, and harmonic with the negative slope in dashed gray. Note that, because of glide symmetry, there is no bandgap at $\beta_{\rm{c}}=3\pi/d$, there is a wide band synchronization, and the two branches have different polarizations (the backward wave is not $z$ polarized). The beamline $\beta_0=\omega/ u_0$ (dotted red) denotes the electron beam with average velocity $u_0$ that synchronizes over a very wide band with the blue branch representing the forward mode with $z$-polarized electric field.}
\label{fig:disp_diagr}
\end{figure}

The dispersion diagram of the waveguide modes, illustrated in Fig. \ref{fig:disp_diagr}, was computed using the eigenmode solver of the commercial software CST Studio Suite. In the eigenmode simulations, the unit cell was composed of perfect electric conductor (PEC) material. In this analysis, we adopt the Brillouin zone (BZ) definition where the fundamental (0th order) BZ corresponds to $-1 < \beta_{\rm{c}} d/\pi < 1$. Consequently, the $n=1$ Floquet harmonic is defined for $1 < \beta_{\rm{c}} d/\pi < 3$, marking the first high-order BZ. Space-harmonic TWTs, such as the serpentine and coupled-cavity-type TWTs are typically designed for operation in the $n=1$ space harmonic.
The small frequency separation between the beam line and the forward branch of the guided wave in Fig. \ref{fig:disp_diagr} is necessary to account for the space-charge effect.

The use of GS in the design effectively suppresses the coupling of space harmonics at the $\beta_{\rm{c}} d = \pi+2\pi n$ points ($n=0, \pm 1, \pm 2, \ldots$), preventing bandgaps from occurring at these values due to the distinct parity exhibited by the field at different harmonics, as demonstrated in \cite{Zvonimir_2024}. Analysis of the field at the beam location reveals the absence of an $E_z$-component in the backward wave (plotted in gray in Fig.\ref{fig:disp_diagr}). Conversely, for the forward branch (blue), the field does exhibit this component, as illustrated in \cite{Castro23}, thereby minimizing the potential for oscillations while allowing interaction with the forward branch. It is important to note that this effect due to GS requires precise alignment of the top and bottom plates. Longitudinal or transverse misalignment will cause bandgaps to open up at the edges of the BZ \cite{Nguyen2014,liu2016characteristics}. Tolerance studies will be necessary to determine what amount of misalignment is acceptable for both manufacturing and TWT operation. 

Using GS, we benefit from low-dispersion in the $\beta_{\rm{c}}-\omega$ relation (i.e. where the group velocity and phase velocity of the guided mode are nearly equal) over a wide frequency range. As a result, interaction with an electron beam can occur within the frequency span of 54.5 GHz to 66.25 GHz, yielding a maximum fractional bandwidth of approximately 20\%. This represents a notably broader bandwidth compared to similar TWT designs, such as the one shown in \cite{Paoloni23}, where a
rectangular WG (non-GS) is presented and the bandwidth is around 8 GHz (spanning from 68 GHz to 76 GHz).


\begin{figure}[htpb]
\centering
\includegraphics[width=0.95\linewidth]{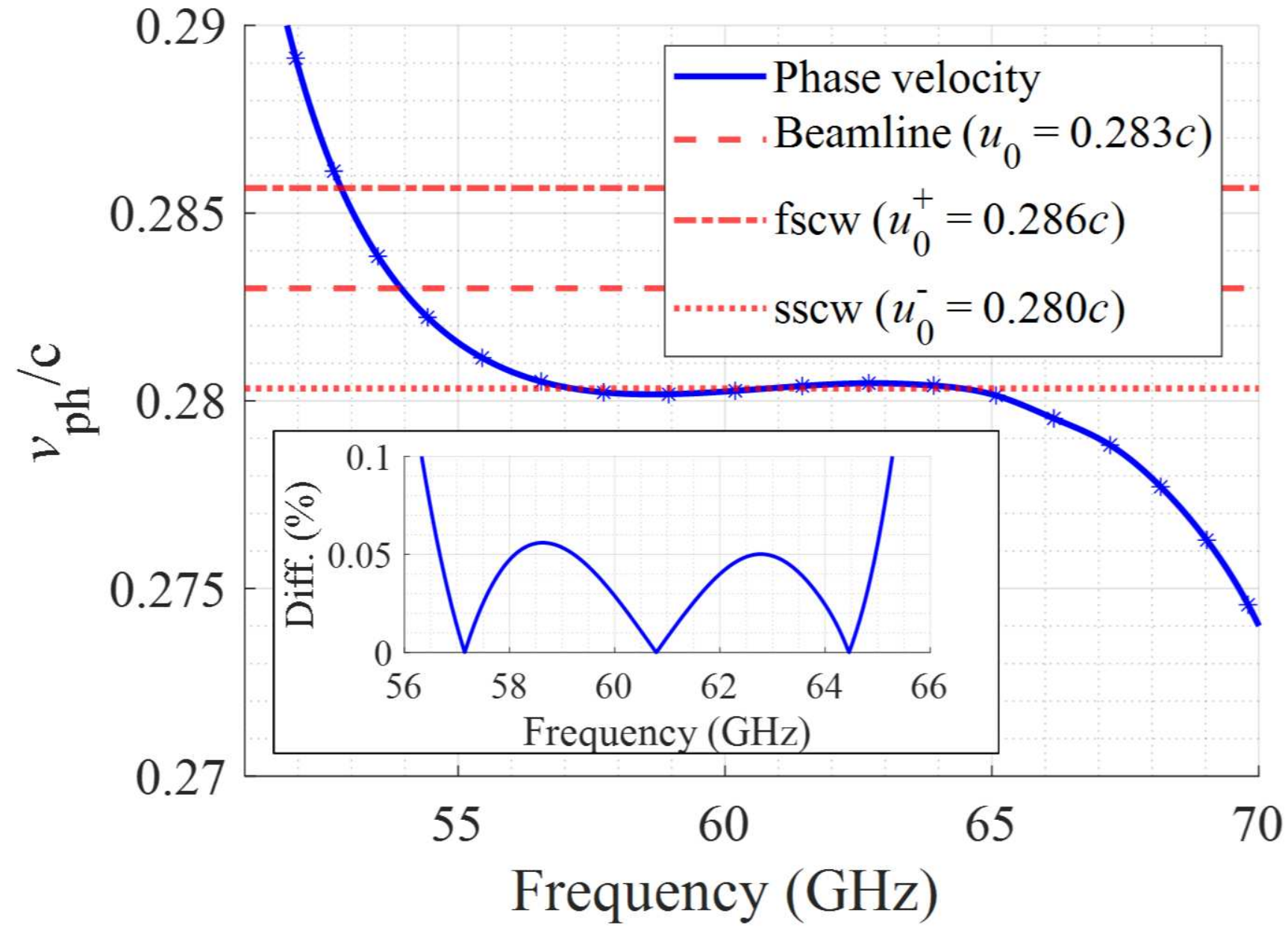}
\caption{Normalized phase velocity of the cold guided wave (blue), and velocities of 
 the electron beam’s
charge waves. The difference between the velocity of the slow space-charge wave (sscw) and the velocity of the guided wave is shown in the inset.}
\label{fig:phase_vel_diff}
\end{figure}

The velocities of the fast and slow space-charge waves are defined as $u_0^\pm = \omega/(\beta_0 \mp \beta_{\rm{q}})$, with $\beta_{\rm{q}} = \omega_{\rm{q}}/u_0$ being the plasma phase constant of the beam traveling with phase velocity $u_0$ and at a reduced angular plasma frequency $\omega_{\rm{q}}$, and $\beta_0=\omega/u_0$. The reduced plasma angular frequency is defined as $\omega_{\rm{q}} = R_{\rm{sc}}\omega_{\rm{p}}$, where $R_{\rm{sc}}$ is the plasma frequency reduction factor, and $\omega_{\rm{p}}$ is the angular plasma frequency of the beam defined as $\omega_{\rm{p}}^2 = I_0u_0/(2V_0A_{\rm{b}}\epsilon_0)$, with $I_0$ being the DC current, $V_0$ the DC beam voltage, $A_{\rm{b}}$ the cross-section area of the beam, and $\epsilon_0$ as the dielectric constant of free space \cite{BranchMihran55}. The velocity of the \textit{slow space-charge wave} (sscw) is $u_0^- =\omega/(\beta_0 + \beta_{\rm{q}})= 0.280c$. The average velocity of the beam was swept in PIC simulations and was selected to obtain the best synchronization (i.e. the maximum possible 3 dB-bandwidth). The plasma frequency reduction factor associated with $u_0^-$ was then empirically determined using the selected average beam velocity and is approximately $R_{\rm{sc}}=0.29$ ($\beta_{\rm{q}} = 44.8$ rad/m). The reduction factor is assumed to be approximately constant over the operating band in Fig. \ref{fig:phase_vel_diff}. The average beam velocity and beam current were chosen to be $u_0 = 0.283c$ and $I_0 = 25$ mA, respectively.  

Fig. \ref{fig:phase_vel_diff} illustrates the normalized phase velocity of the guided wave and its synchronization with the sscw velocity in the electron beam. The inset plot highlights the percentage difference between the sscw velocity and the phase velocity of the guided wave, defined as $\mathrm{Diff} \% = (|u_0^- - v_{\rm{ph}}|/u_0^-)100$, showing that this difference remains below $0.1\%$ over a band of approximately 9 GHz. This result suggests a promising synchronization between the phase velocity of the guided wave and the sscw, which could contribute to achieving a uniform gain profile across a broad spectrum. 

\subsection{Interaction Impedance}

The interaction impedance serves as a metric to measure the coupling between the RF wave and the electron beam, essentially quantifying the $E_z$ field component of a guided mode at the location where the electron beam is expected to flow \cite{pierce-twt50}. For devices operating in the $n=1$ space harmonic, it is formally found as \cite{gewartowski65CH10}

\begin{equation}
    Z_{\mathrm{P}} = \frac{|E_{z,1}|^2}{2\beta_1^2P}     \label{eqn:interaction_impedance}
\end{equation}

\noindent where $\beta_1 = \beta_{\rm{c}} + 2\pi/d$ is the phase constant of the $1$st Floquet harmonic and $P$ is the  time-average electromagnetic power flow for the guided mode \cite{pierce-twt50}. The quantity $E_{z,1}$ is the phasor representing the longitudinal component of the $1$st spatial harmonic of the electric field for the guided mode of interest. 

\begin{figure}[htpb]
\centering
\includegraphics[width=0.95\linewidth]{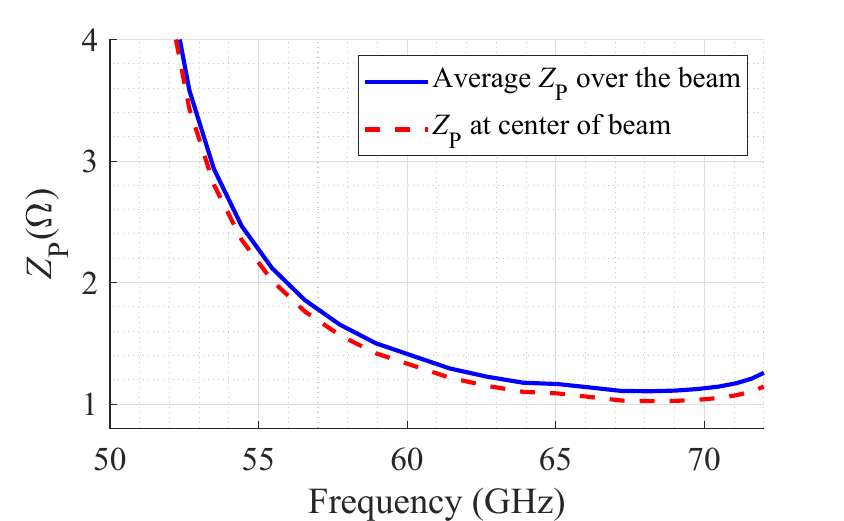}
\caption{Interaction impedance at the center of the electron beam (dashed red) and the average value over the cross-section of the electron beam (solid blue).}
\label{fig:Zp_Average}
\end{figure}

In Fig. \ref{fig:Zp_Average}, we show the interaction impedance as a function of the frequency for the positive-slope (i.e. forward propagating) branch of the modal dispersion diagram shown in Fig. \ref{fig:disp_diagr}. The dashed red line represents the interaction impedance at the center of the electron beam's circular cross-section, while the solid blue line corresponds to the average interaction impedance over the cross-section of the beam. Notably, the average interaction impedance, which is approximately $Z_{\mathrm{P}} \approx 1$ $\Omega$ at the center operating frequency, is primarily influenced by the transverse separation (i.e. in $x$- or $y$-directions in Fig. \ref{fig:unit_cell}) between the inner pillars. In the case of a BT glide structure, the interaction impedance tends to decrease as the gap between the pillars widens \cite{Castro23}.

\begin{figure}[htpb]
\centering
\includegraphics[width=0.95\linewidth]{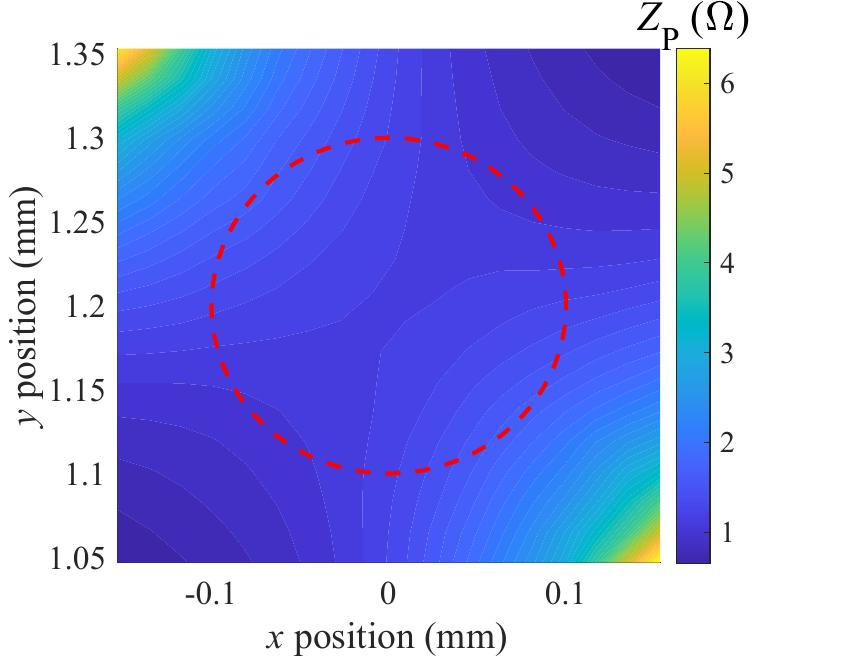}
\caption{Interaction impedance over the cross-section of the tunnel when the frequency is $60$ GHz and the normalized phase constant is $\beta_{\rm{c}} d/\pi = 2.75$. The region inside the dashed red circle corresponds to the electron beam region.} 
\label{fig:Zp_Surface}
\end{figure}

Furthermore, Fig. \ref{fig:Zp_Surface} illustrates the distribution of the interaction impedance within the region situated between the two inner pillars of the unit cell when the frequency is $60$ GHz. The figure highlights the nearly uniform distribution of the interaction impedance across the entire cross-sectional area of the beam. The bright spots where the interaction impedance is high correspond to the corners of the inner pillars. This uniformity of interaction impedance across the beam cross-section is crucial as it allows the guided mode of the SWS to efficiently exchange energy with the synchronized electron beam.

\section{TWT Gain and Bandwidth Analysis}

We conducted several three-dimensional particle-in-cell (3-D PIC) simulations using the PIC solver of CST Studio Suite. To verify that the TWT will not oscillate in the small-signal regime, a 3-D PIC simulation without RF excitation at Port 1 or Port 2 (i.e. a zero-drive stability test) was conducted first. The electron beam has a velocity $u_0 = 0.283c$, a dc current $I_0 = 25$ mA, and is confined by a magnetic field of approximately $B_z = 0.55$ T, which is five times the Brillouin flux density for this electron beam \cite{Gilmour11}. The beam is also modeled with a beam rise time of $t_r = 0.1$ ns, during which the electron beam current increases smoothly from zero to $I_0$. The output power  ${\it P}_o$ at Port 2, at the collector-end of the TWT, remained below 5 nW after 70 ns in zero-drive simulations. This power level is considered negligible, especially when compared to the expected input power of the RF signal that will be later introduced at Port 1 of the TWT, suggesting that the structure will not exhibit exponentially growing oscillations for the selected beam parameters and dimensions.

\begin{figure}[htpb]
\centering
\includegraphics[width=0.95\linewidth]{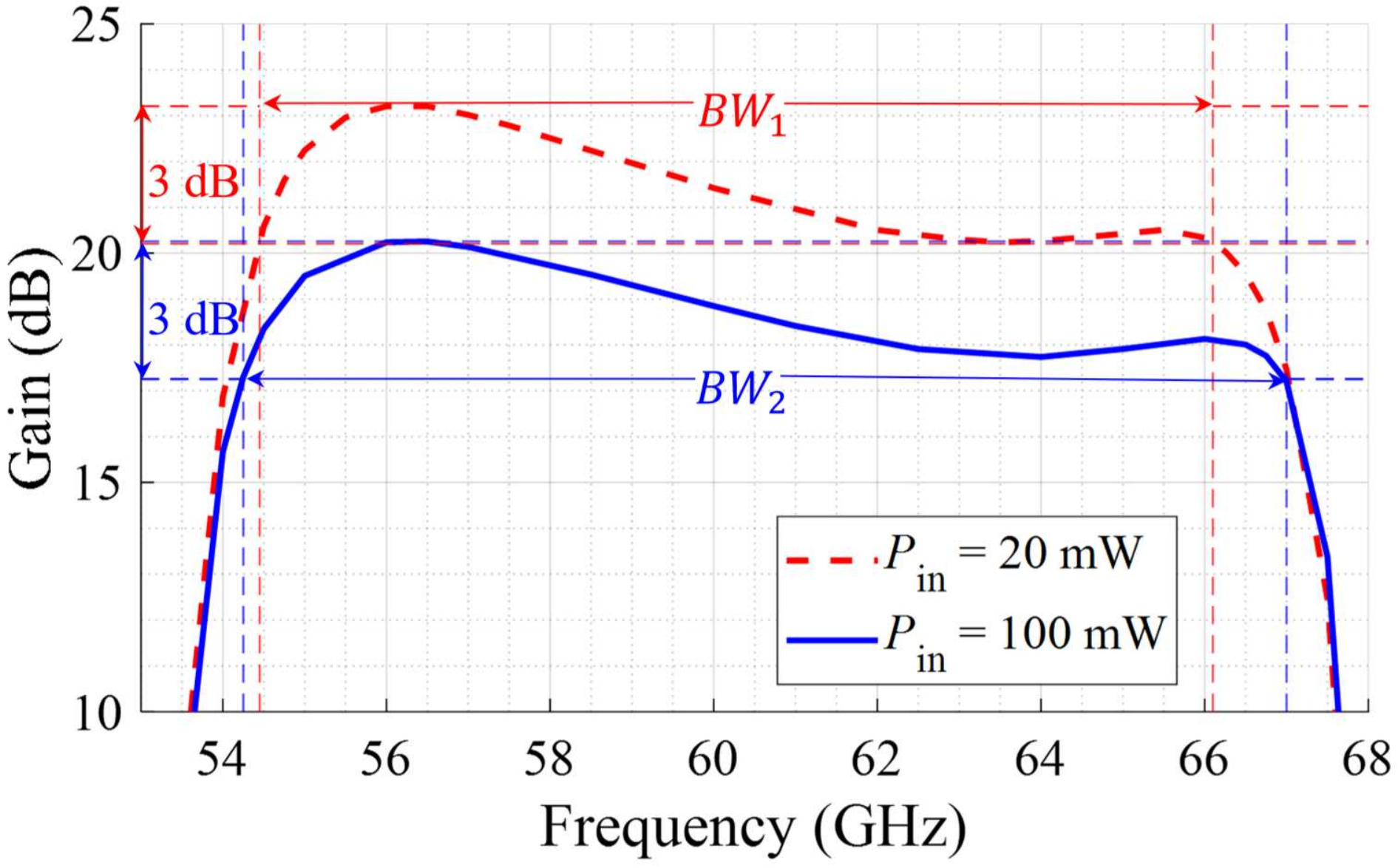}
\caption{Gain of the TWT with $N_c=95$, $u_0 = 0.283c$  and $I_0 = 25$ mA, for two different values of input power, $P_{\mathrm{in}}$. Note that a wide 3-dB gain bandwidth is enabled by glide symmetry.} 
\label{fig:gain_vs_freq}
\end{figure}

The TWT design was studied with full-wave PIC simulations to obtain the following gain plots. Fig. \ref{fig:gain_vs_freq} shows the gain and 3-dB bandwidth for two different values of input power applied to Port 1 of the TWT, 20 mW, and 100 mW. The simulation setup uses approximately 88 million mesh cells and 7.1 million particles, the electron beam velocity and current are identical to those used in the zero-drive stability analysis (i.e. $u_0 = 0.283c$ and $I_0 = 25$ mA). The gain (in decibels) is defined as $G = FFT_o-FFT_i$, where $FFT_o$ is the magnitude of the spectrum of the output signal and $FFT_i$ is the magnitude of the spectrum of the input signal. For an input power of $20$ mW, the maximum gain is 23.23 dB at 56.5 GHz, and the 3-dB gain bandwidth, $BW_1$, is 11.75 GHz, spanning from 54.5 GHz to 66.25 GHz, which is equivalent to a fractional bandwidth of around 19.5\%. With an input power of $100$ mW, the maximum gain is 20.25 dB at the same frequency (56.5 GHz), and the 3-dB gain bandwidth, $BW_2$, is around 12.75 GHz, spanning from 54.25 GHz to 67 GHz, resulting in a fractional bandwidth of about 21\%. For these two input powers, there is a difference of approximately $1$ GHz in the 3-dB gain bandwidth and the difference in the fractional bandwidth is 1.5\%. 

\begin{figure}[htpb]
\centering
\includegraphics[width=1\linewidth]{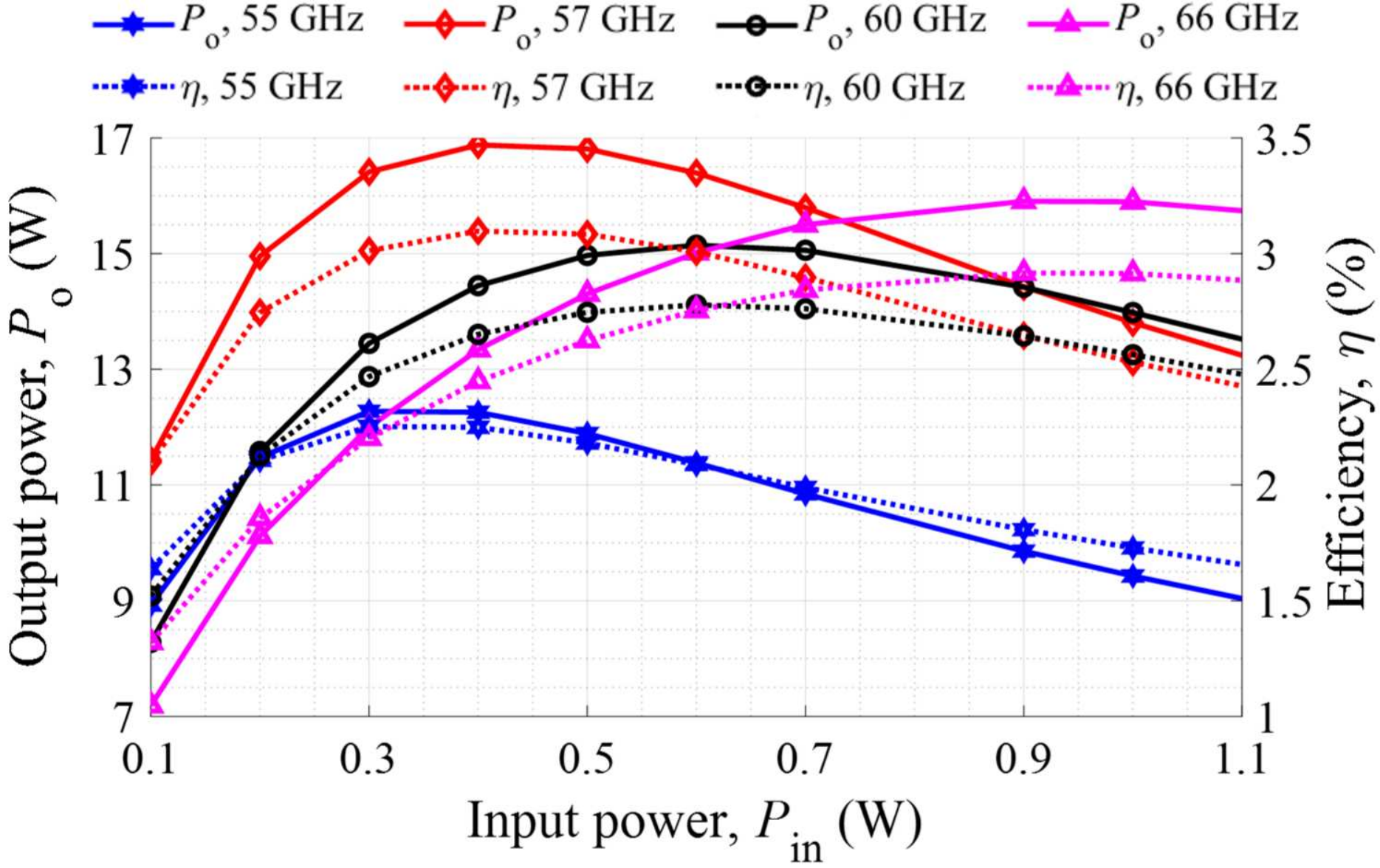}
\caption{Output power and efficiency of the TWT for different values of input power at four different frequencies, assuming  $u_0 = 0.283c$ and $I_0 = 25$ mA. }
\label{fig:Pout_Efficiency_BT-G_TWT_v2}
\end{figure}


The proposed design exhibits a remarkably wider fractional bandwidth in comparison with other designs with similar features found in the literature, also considering that only one uniform interaction section (i.e., no severs) is included. In \cite{Paoloni23}, for example, a rectangular WG with an operational band around 71-76 GHz (fractional bandwidth of around 7\%) and two stages is presented. In \cite{Liao15V_band_TPS}, a serpentine WG with bandwidth of 5 GHz and central frequency around 60 GHz is described (fractional bandwidth of around 8\%). In \cite{zhuge2015design}, another serpentine WG is shown, with a bandwidth of 5 GHz around 55-60 GHz (fractional bandwidth of around 9\%). Also, in \cite{Chong10}, several commercial and military helix TWT designs are presented, with bandwidths ranging from 1 GHz to 15 GHz in Ka- and Q-band frequency ranges (fractional bandwidths as high as 46\%).





Figure \ref{fig:Pout_Efficiency_BT-G_TWT_v2} shows the DC-RF conversion efficiency performance of the structure. The highest saturation power is 16.9 W at around 57 GHz with an input drive of 0.4 W, and the maximum basic efficiency is approximately 3.1\%, defined as $\eta = P_o/P_{\mathrm{i,total}}$, where $P_{\mathrm{i,total}} = P_{\mathrm{b}}+{\it P}_{\mathrm{in}}$, with $P_{\mathrm{b}} = I_0V_0$ being the power of the beam, and ${\it P}_{\mathrm{in}}$ being the input RF power at Port 1.

\begin{figure}[htpb]
\centering
\includegraphics[width=0.95\linewidth]{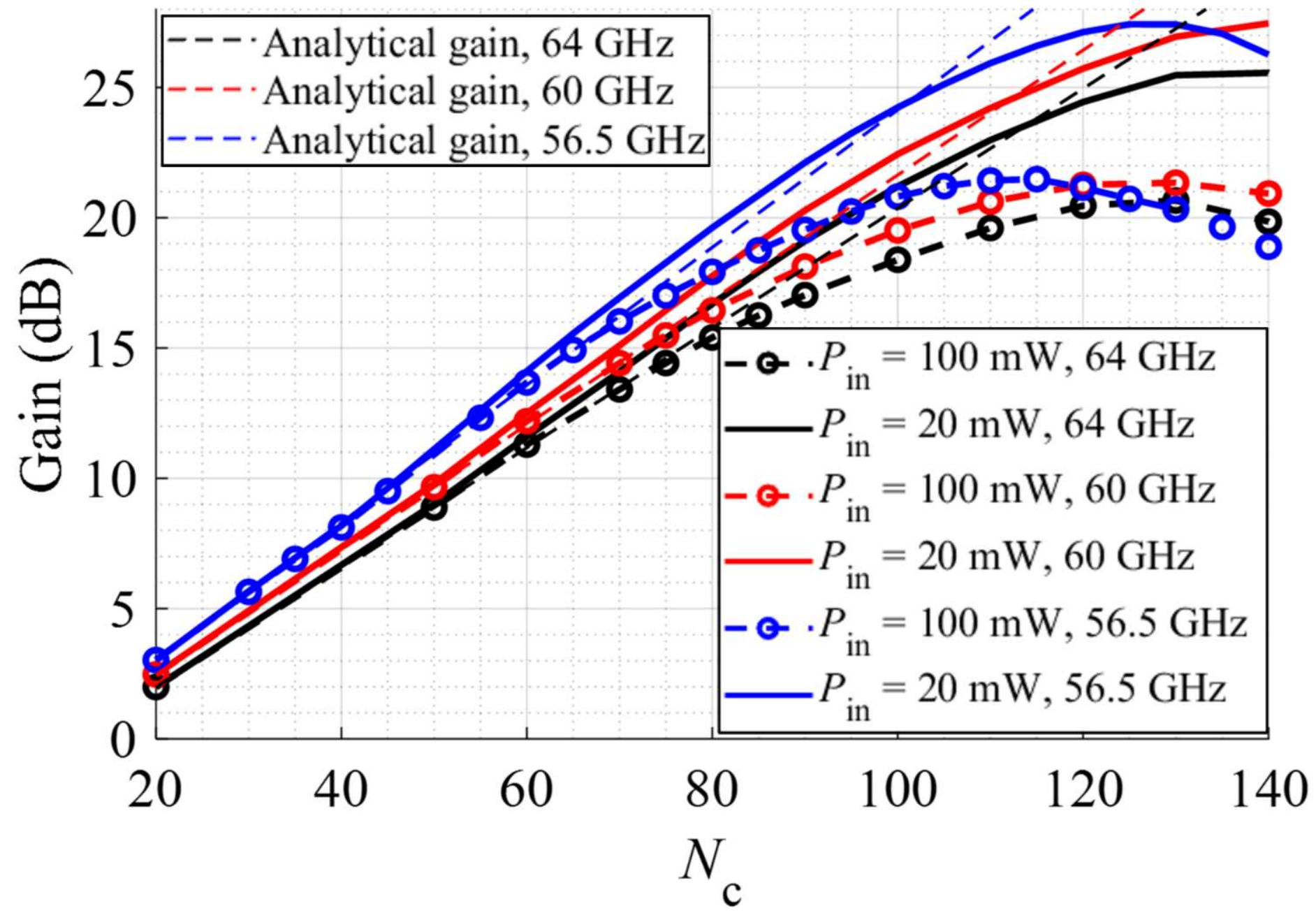}
\caption{Gain versus numbers of unit cells $N_{\rm{c}}$, with $u_0 = 0.283c$ and $I_0 = 25$ mA. Three different frequencies and two different values for the input power, ${\it P}_{\mathrm{in}}$, are considered for comparison of PIC simulations with the analytical Pierce model (dashed).}
\label{fig:gain_vs_N_plot}
\end{figure}

Figure \ref{fig:gain_vs_N_plot} shows the gain performance for different lengths of the interaction section in the structure, represented by various numbers of unit cells ${\it N}_c$. The data is shown for three frequencies: 56.5 GHz (frequency of maximum gain),  60 GHz, and 64 GHz, and two values of RF input power: $20$ mW and $100$ mW. The dashed lines represent the theoretical small-signal gain, presented in \cite{pierce-twt50}, defined as $G_{dB} = A + BCN_{\lambda}$. Here,  $N_{\lambda}=\beta_0N_{c}d/(2\pi)$ is the number of electronic wavelengths of the circuit, with $\beta_0$ being the propagation constant of the beam,  $C = (Z_\mathrm{P}I_0/(4V_0))^{1/3}$ is the dimensionless Pierce gain parameter, and $A$ and $B$ (in dB) are parameters that account for the losses and space-charge effect, fitted to the PIC gain curves in the small signal regime. The Pierce gain parameter can also be expressed in terms of the characteristic impedance of the SWS and the beam-wave coupling coefficient, as was done in \cite{rouhi2021exceptional,marosi2024small}. For this analysis, perfect synchronism is assumed for the frequency $f_{\rm{m}} = 56.5$ GHz, as it is the frequency at which
the maximum gain is obtained, which implies that $\beta_0 = \beta_{\rm{c}}(f_{\rm{m}})$. The fitted values (in dB) are $(A,B)=(-2.25, 24.9)$ at 56.5 GHz, $(A,B)=(-2.32, 25.5)$ at 60 GHz, and $(A,B)=(-2.6, 25.9)$ at 64 GHz. As expected, good agreement between the PIC curves and the theoretical Pierce power gain is evident for the small signal regime region, and the gain saturation is higher when a smaller input power is considered.


\section{Conclusion}
We have introduced the GSDC-GW that combines the benefits of GS and EBG, allowing for wide-band operation without requiring a conductive connection between the upper and lower waveguide plates. We evaluated the performance of this structure using particle-in-cell (PIC) simulations, which revealed an impressive GS-induced 12.75-GHz 3-dB gain bandwidth, spanning from 54.25 GHz to 67 GHz, along with a maximum gain of 20.25 dB, which is equivalent to a wide fractional bandwidth of around 21\% that is significantly wider than typical bandwidths of other space-harmonic TWTs with similar features found in the literature. The GS fundamental principles highlighted in this paper set up guidelines for promising candidates among the various SWS topologies suitable for very wide band TWTs in millimeter-wave frequencies.

\section*{Acknowledgment}


The authors are thankful to DS SIMULIA for providing CST Studio Suite that was instrumental in this study. 

\appendix 
\section*{Geometry Description} \label{Appendix:A}

\begin{figure}[htpb]
\centering
\includegraphics[width=0.95\linewidth]{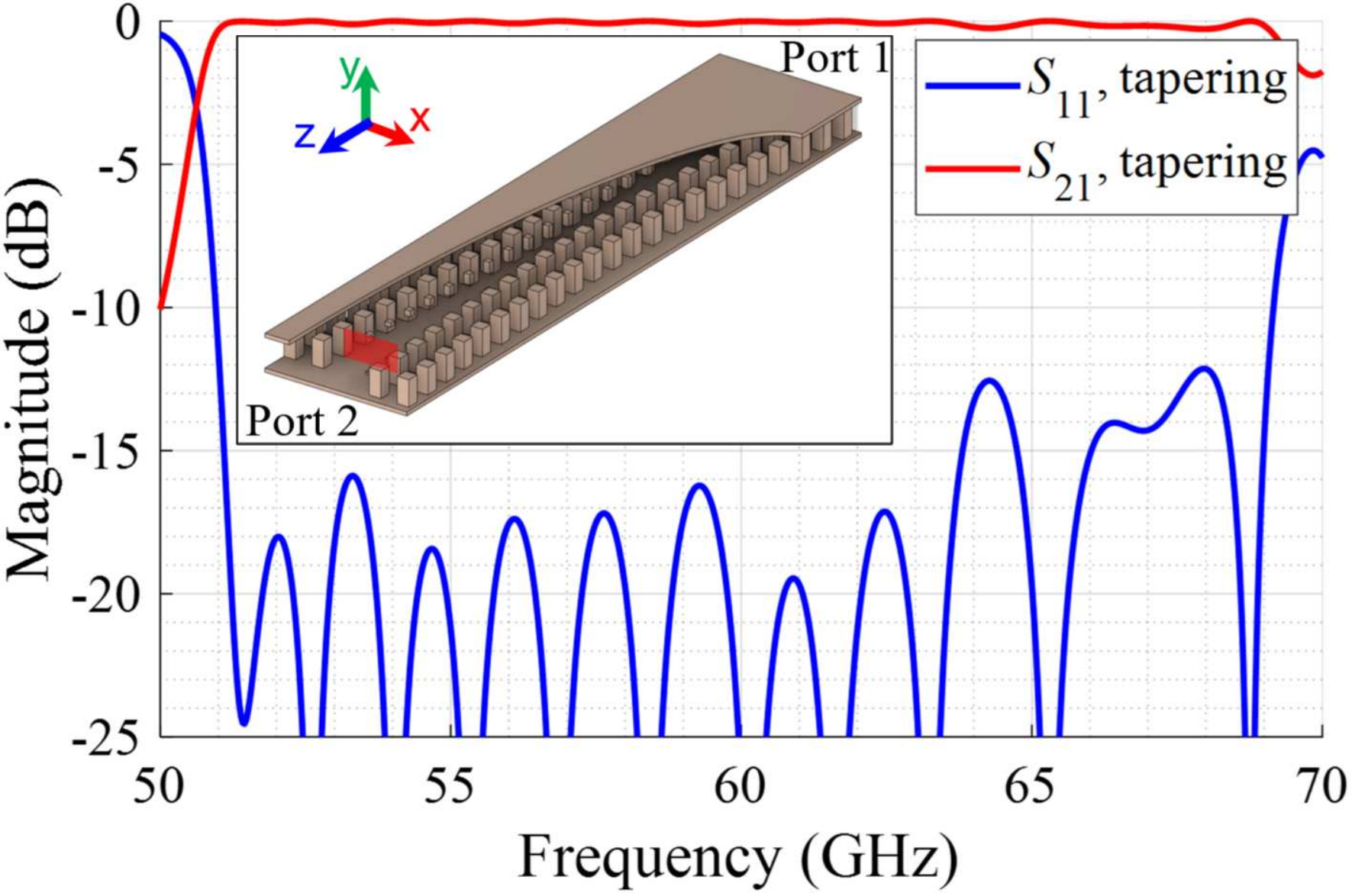}
\caption{Scattering parameters of GS waveguide made of an interaction section of 10 unit cells and two nonlinear tapering transitions consisting of 11 cells each. All tapering cells are GS in order to transition the guided modes of the SWS into the TE10 mode of a rectangular waveguide with minimal reflection.}
\label{fig:Spar_tapering}
\end{figure}

To evaluate the effectiveness of the 11-step tapering sections designed for coupling the SWS with the standard WR15 waveguide, a full-wave simulation was conducted. 
This simulation involved incorporating a $N_c=10$ unit cell interaction section and attaching the tapering sections to each end, still following the GS guidelines, and simulating $S_{11}$ and  $S_{21}$, illustrated in Fig. \ref{fig:Spar_tapering}. 
The transition of inner pillars follows a cosine profile variation in height, while the transition of the outer pillar uses a cosine squared profile in the position along $x$. The inner and outer pillars are still attached to the top and bottom plates in the taper section following GS guidelines, just as in the interaction section. The same occurs in the bend sections, but there are no inner pillars in the bend sections. 
The final design shows that $S_{11}$ remains below $-13$ dB, while $S_{21}$ stays above $-0.1$ dB, across the frequency range spanning from 50 GHz to 70 GHz.

\begin{figure}[htpb]
\centering
\includegraphics[width=0.95\linewidth]{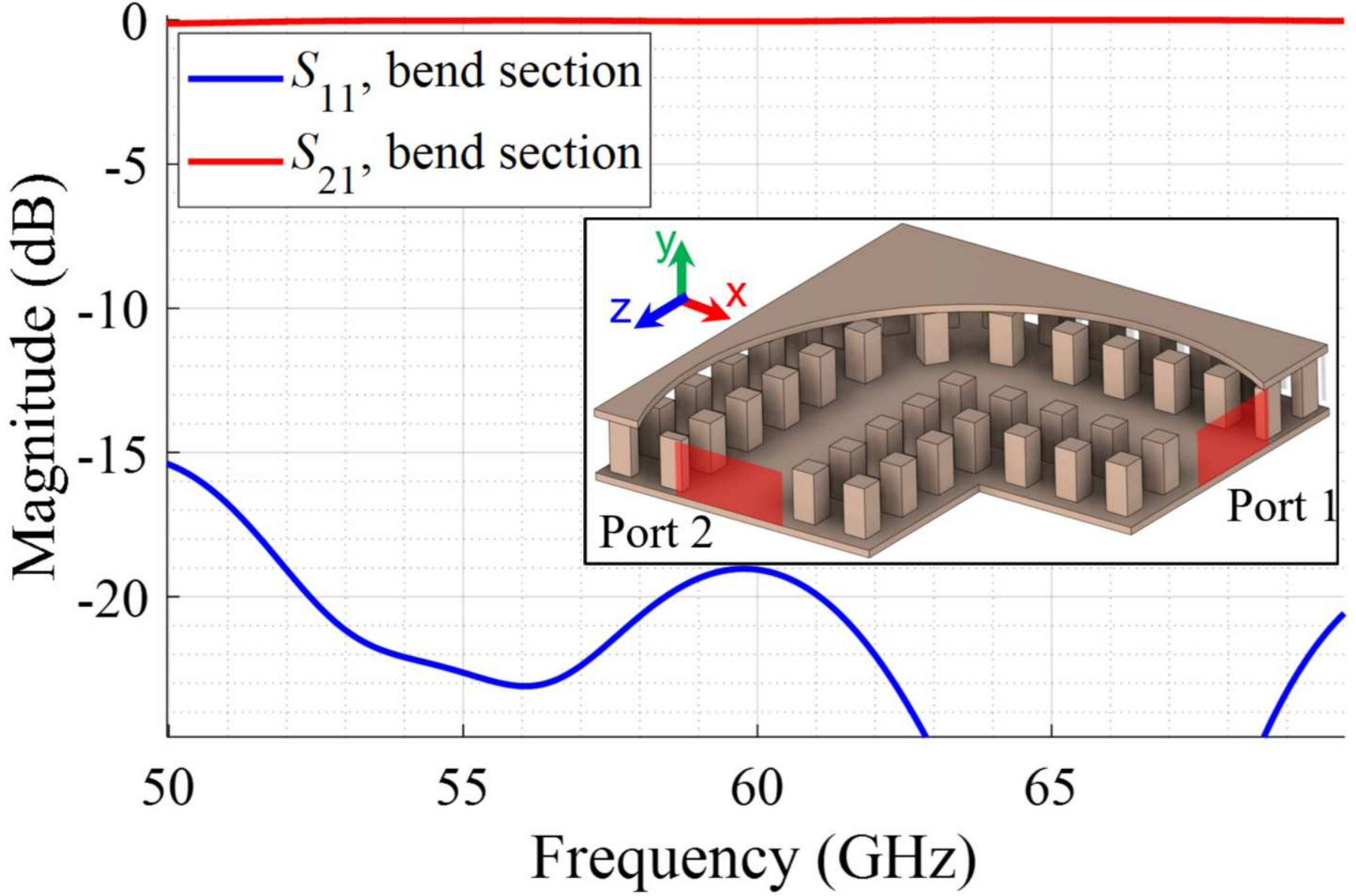}
\caption{Scattering parameters of the bend employed at each end of the TWT. Such bends are designed to be connected with a standard WR15 waveguide.}
\label{fig:Spar_bend}
\end{figure}

A similar simulation setup was employed to analyze the behavior of the bend sections, and the results are visualized in Fig. \ref{fig:Spar_bend}. Within the frequency spectrum spanning 50 GHz to 70 GHz, it is evident that $S_{11}$ consistently remains below $-15$ dB, while $S_{21}$ maintains values above $-0.3$ dB.

\begin{figure}[htpb]
\centering
\includegraphics[width=0.95\linewidth]{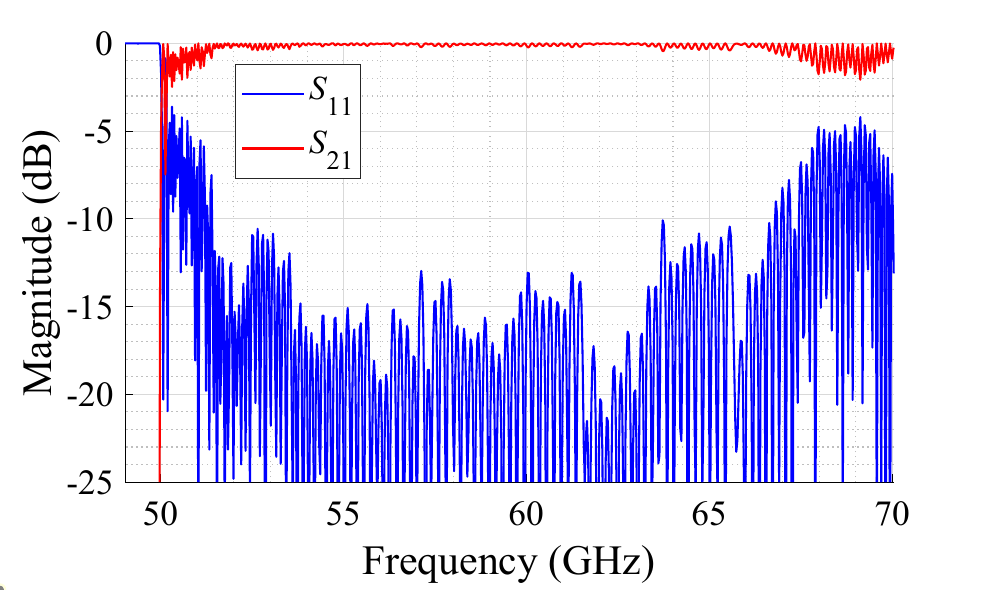}
\caption{Scattering parameters of the full structure with 100 unit cells.}
\label{fig:Spar_FullStr}
\end{figure}

The S-parameters of a DCW circuit with $N_c=100$ unit cells in the interaction section, 11 unit cells in each coupler, and bend sections at both ends are shown in Fig. \ref{fig:Spar_FullStr}. Reflection is below $-10$ dB in the band between 51.5 GHz and 66.5 GHz, indicating a reduced risk of oscillations for high gain values approaching $20$ dB. The number of unit cells in the interaction section in this simulation is slightly greater than the length of the interaction section in the final design (five more unit cells) to ensure that the stability achieved is reliable.

\bibliographystyle{IEEEtran}
\bibliography{TED_DoubleCorrugatedWaveguide}

\end{document}